%Paper: astro-ph/9506129
%From: Ed Turner <elt@astro.Princeton.EDU>
%Date: Tue, 27 Jun 95 12:52:09 EDT

%%                      	JNL.TEX
%%
%%                This is JNL.TEX Version 0.3 as of 6/12/85.
%%
%%	This is a set of TeX 82 macros designed to produce scientific
%%	papers with a minimum of fuss and using as much of plain.tex as
%%	possible.  The user need only know what is in the TeXbook, and
%%	the macros under ``user definitions'' below.  Also, the user
%%	definitions are intended to be as simple as possible, so that
%%	the user may change them as desired.

%%
%%  Font definitions suitable for the IMAGEN (Written by Tony Kennedy)compare
%%with others.
%%

%  Define a whole menagerie of pseudo-12pt fonts

\font\twelverm=cmr10 scaled 1200    \font\twelvei=cmmi10 scaled 1200
\font\twelvesy=cmsy10 scaled 1200   \font\twelveex=cmex10 scaled 1200
\font\twelvebf=cmbx10 scaled 1200   \font\twelvesl=cmsl10 scaled 1200
\font\twelvett=cmtt10 scaled 1200   \font\twelveit=cmti10 scaled 1200

\skewchar\twelvei='177   \skewchar\twelvesy='60

%  Define \...point macros to change fonts and spacings consistently

\def\twelvepoint{\normalbaselineskip=12.4pt
  \abovedisplayskip 12.4pt plus 3pt minus 9pt
  \belowdisplayskip 12.4pt plus 3pt minus 9pt
  \abovedisplayshortskip 0pt plus 3pt
  \belowdisplayshortskip 7.2pt plus 3pt minus 4pt
  \smallskipamount=3.6pt plus1.2pt minus1.2pt
  \medskipamount=7.2pt plus2.4pt minus2.4pt
  \bigskipamount=14.4pt plus4.8pt minus4.8pt
  \def\rm{\fam0\twelverm}          \def\it{\fam\itfam\twelveit}%
  \def\sl{\fam\slfam\twelvesl}     \def\bf{\fam\bffam\twelvebf}%
  \def\mit{\fam 1}                 \def\cal{\fam 2}%
  \def\tt{\twelvett}
  \def\nullspace{\nulldelimiterspace=0pt \mathsurround=0pt }
  \def\big##1{{\hbox{$\left##1\vbox to 10.2pt{}\right.\nullspace$}}}
  \def\Big##1{{\hbox{$\left##1\vbox to 13.8pt{}\right.\nullspace$}}}
  \def\bigg##1{{\hbox{$\left##1\vbox to 17.4pt{}\right.\nullspace$}}}
  \def\Bigg##1{{\hbox{$\left##1\vbox to 21.0pt{}\right.\nullspace$}}}
  \textfont0=\twelverm   \scriptfont0=\tenrm   \scriptscriptfont0=\sevenrm
  \textfont1=\twelvei    \scriptfont1=\teni    \scriptscriptfont1=\seveni
  \textfont2=\twelvesy   \scriptfont2=\tensy   \scriptscriptfont2=\sevensy
  \textfont3=\twelveex   \scriptfont3=\twelveex  \scriptscriptfont3=\twelveex
  \textfont\itfam=\twelveit
  \textfont\slfam=\twelvesl
  \textfont\bffam=\twelvebf \scriptfont\bffam=\tenbf
  \scriptscriptfont\bffam=\sevenbf
  \normalbaselines\rm}

%	tenpoint

%%
%%	Various internal macros
%%

\def\beginlinemode{\endmode
  \begingroup\parskip=0pt \obeylines\def\\{\par}\def\endmode{\par\endgroup}}
\def\beginparmode{\endmode
  \begingroup \def\endmode{\par\endgroup}}
\let\endmode=\par
{\obeylines\gdef\
{}}
\def\singlespace{\baselineskip=\normalbaselineskip}

\def\oneandahalfspace{\baselineskip=\normalbaselineskip
  \multiply\baselineskip by 3 \divide\baselineskip by 2}
\def\doublespace{\baselineskip=\normalbaselineskip \multiply\baselineskip by 2}

\newcount\firstpageno
\firstpageno=2
%% FOLLOWING LINE CANNOT BE BROKEN BEFORE 80 CHAR
\footline={\ifnum\pageno<\firstpageno{\hfil}\else{\hfil\twelverm\folio\hfil}\fi}
\let\rawfootnote=\footnote		% We must set the footnote style
\def\footnote#1#2{{\rm\singlespace\parindent=0pt\rawfootnote{#1}{#2}}}
\def\raggedcenter{\leftskip=4em plus 12em \rightskip=\leftskip
  \parindent=0pt \parfillskip=0pt \spaceskip=.3333em \xspaceskip=.5em
  \pretolerance=9999 \tolerance=9999
  \hyphenpenalty=9999 \exhyphenpenalty=9999 }
\def\dateline{\rightline{\ifcase\month\or
  January\or February\or March\or April\or May\or June\or
  July\or August\or September\or October\or November\or December\fi
  \space\number\year}}
\def\received{\vskip 3pt plus 0.2fill
 \centerline{\sl (Received\space\ifcase\month\or
  January\or February\or March\or April\or May\or June\or
  July\or August\or September\or October\or November\or December\fi
  \qquad, \number\year)}}

%%
%%	Page layout, margins, font and spacing (feel free to change)
%%

\hsize=6.5truein
%\hoffset=1truein
\vsize=8.9truein
%\voffset=1truein
\parskip=\medskipamount
\twelvepoint		% selects twelvepoint fonts (cf. \tenpoint)
\doublespace		% selects double spacing for main part of paper (cf.
			%	\singlespace, \oneandahalfspace)
\overfullrule=0pt	% delete the nasty little black boxes for overfull box

%%
%%	The user definitions for major parts of a paper (feel free to change)
%%

	% Preprint number at upper right of title page

\def\title			%  Title on title page
  {\null\vskip 3pt plus 0.2fill
   \beginlinemode \doublespace \raggedcenter \bf}

\def\author			%  Author(s) name(s)  on title page
  {\vskip 3pt plus 0.2fill \beginlinemode
   \singlespace \raggedcenter}

\def\affil			% Affiliations (can intermix with \author)
  {\vskip 3pt plus 0.1fill \beginlinemode
   \oneandahalfspace \raggedcenter \sl}

\def\abstract			% Begin abstract
  {\vskip 3pt plus 0.3fill \beginparmode
   \doublespace \narrower ABSTRACT: }

\def\endtitlepage		% End title page, begin body of paper
  {\endpage			% 	This subsumes \body
   \body}

\def\body			% Begin text body;  can be used to end
  {\beginparmode}		% \title, \author, \affil, \abstract,
				% \reference, or \figurecaption modes

\def\head#1{			% Head;  NOTE enclose the text in {}
  \filbreak\vskip 0.5truein	%  e.g., \head{I. Introduction}
  {\immediate\write16{#1}
   \raggedcenter \uppercase{#1}\par}
   \nobreak\vskip 0.25truein\nobreak}

\def\refto#1{$^{#1}$}		% For references in text as superscript

\def\references			% Begin references -- basic format is Phys Rev
  {\head{References}		% I.e., volume, page, year (space after commas).
   \beginparmode
   \frenchspacing \parindent=0pt \leftskip=1truecm
   \parskip=8pt plus 3pt \everypar{\hangindent=\parindent}}

\gdef\refis#1{\indent\hbox to 0pt{\hss#1.~}}	% Ref list numbers.

\gdef\journal#1, #2, #3, 1#4#5#6{		% Journal reference.  Comma sets
    {\sl #1~}{\bf #2}, #3, (1#4#5#6)}		% off: name, vol, page, year

\gdef\journ2 #1, #2, #3, 1#4#5#6{		% Journal reference.  Comma sets
    {\sl #1~}{\bf #2}: #3, (1#4#5#6)}		% off: name, vol, page, year
                                     		% Colon inserted after volume #

\def\refstylenp{		% Nucl Phys(or Phys Lett) ref style: V, Y, P
  \gdef\refto##1{ [##1]}				% Reference in text []
  \gdef\refis##1{\indent\hbox to 0pt{\hss##1)~}}	% Ref list numbers)
  \gdef\journal##1, ##2, ##3, ##4 {			% Journal reference
     {\sl ##1~}{\bf ##2~}(##3) ##4 }}

\def\refstyleprnp{		% Input like pr, output like np!!
  \gdef\refto##1{ [##1]}				% Reference in text []
  \gdef\refis##1{\indent\hbox to 0pt{\hss##1)~}}	% Ref list numbers)
  \gdef\journal##1, ##2, ##3, 1##4##5##6{		% Journal reference
    {\sl ##1~}{\bf ##2~}(1##4##5##6) ##3}}

\def\figurecaptions		% Begin figure captions
  {\endpage
   \beginparmode
   \head{Figure Captions}
}

\def\endpage			%  Eject a page
  {\vfill\eject}

\def\endpaper			%  Ways to say goodbye
  {\endmode\vfill\supereject}

%%
%%	Various little user definitions
%%

\def\ref#1{Ref. #1}			% 	for inline references
			% 	ditto

		% For citation of equation numbers
	%	ditto
			%	ditto
			%	ditto
			%	ditto
			%	ditto
\def\frac#1#2{{\textstyle #1 \over \textstyle #2}}

\def\sla{\raise.15ex\hbox{$/$}\kern-.57em}
\def\leaderfill{\leaders\hbox to 1em{\hss.\hss}\hfill}
\def\twiddle{\lower.9ex\rlap{$\kern-.1em\scriptstyle\sim$}}
\def\bigtwiddle{\lower1.ex\rlap{$\sim$}}
\def\gtwid{\mathrel{\raise.3ex\hbox{$>$\kern-.75em\lower1ex\hbox{$\sim$}}}}
\def\ltwid{\mathrel{\raise.3ex\hbox{$<$\kern-.75em\lower1ex\hbox{$\sim$}}}}
\def\square{\kern1pt\vbox{\hrule height 1.2pt\hbox{\vrule width 1.2pt\hskip 3pt
   \vbox{\vskip 6pt}\hskip 3pt\vrule width 0.6pt}\hrule height 0.6pt}\kern1pt}

\def\doublespace{\baselineskip=20pt plus 2pt\lineskip=3pt minus
1pt\lineskiplimit=2pt}

\def\singlespace{\normalbaselines}
\def\lsim{\mathrel{\mathpalette\@versim<}}
\def\gsim{\mathrel{\mathpalette\@versim>}}
% \magnification=\magstep1

\bigskip
\centerline{\bf QSO's from Galaxy Collisions with Naked Black Holes}

\centerline{\bf M. Fukugita}

\centerline{\it Institute for Advanced Study, Princeton, NJ 08540, USA \&}

\vskip-2mm
\centerline{\it Yukawa Institute, Kyoto University, Kyoto 606, Japan}

\centerline{\bf E. L. Turner}

\centerline{\it Princeton University Observatory, Princeton, NJ 08544, USA}
\bigskip

%ELT draft as of 6/9/95 2pm JST

{\bf
In the now well established conventional view (see Rees [1]
and references therein), quasi-stellar objects (QSOs) and related
active galactic nuclei (AGN) phenomena are
explained as the result of accretion of
plasma onto giant black holes which are postulated to form via
gravitational collapse of the high density regions in the centers of
massive host galaxies.  This model is supported by a wide variety of indirect
evidence and seems quite likely
to apply at least to some observed AGN phenomena.  However, one surprising
set of
new Hubble Space Telescope (HST) observations [2-4] directly
challenges the conventional model, and the
well known evolution of the QSO population raises some
additional, though not widely recognized, difficulties.
We propose here an alternative possibility: the Universe
contains a substantial independent population of super-massive black holes, and
QSO's are a phenomenon that occurs due to their collisions with galaxies
or gas clouds in the intergalactic medium (IGM). This hypothesis
would naturally explain why the QSO population declines very rapidly
towards low redshift, as well as the new HST data.	}

The recent direct observation which calls the traditional model into
question is the result of attempts to image the host galaxies of low
redshift QSO's using HST.  So far, 20 such systems have been imaged
deeply with WFPC2 (Bahcall, Kirhakos \& Schneider [2-4]).
A few of the images show the expected
normal, giant galaxy with the QSO shining from its nucleus;
however, the remainder show a somewhat bewildering array of different
local environments for the QSO activity:  in some cases the QSO is
positioned somewhere in the midst of what appears to be a system of
galaxies in collision but not associated with any obvious galaxy nucleus.
In many cases, there are a few dwarf
galaxies within several kpc of the QSO but none detectable
(to limits well below $L^*$)
directly associated with it.  There are even cases in which the QSO
seems to have no particular association with any visible galaxy, aside
from being a part of some possible loose galaxy group.  In any case, the
data clearly demonstrate that a major fraction of at least low redshift
QSO's do not conform to the most straightforward predictions of the
conventional scenario.

The difficulties for the standard scenario raised by observed
QSO population evolution (see Ref. [5])
are neither so direct nor so clearly recognized,
but they may also represent important clues.  There are in fact three such
puzzles.  The first is that at $z>4$, when the Universe was less than
10\% of its present age (for $\Omega_o=1$), the most distant QSO's we have
so far located were as luminous and roughly as numerous as those present
at any later epoch and far brighter and more common than they are at the
present epoch [6,7].  Moreover, this population of
objects may well be present
at even higher redshifts; they are difficult to locate not because they
are faint (some $z>4$ QSO's are brighter than 18th magnitude!) but because
the comoving volume per redshift interval is decreasing
or increasing less rapidly than at low redshift (depending on the
cosmological model) and the QSO
light is rapidly shifting into the near IR.
The second problem for the conventional model is to explain
why the giant black hole remnants of the QSO's which were so luminous at
$z\sim2$ are so dark and inactive at the present, despite the presence of
a dense ISM and stellar population in the nuclei of many giant galaxies [8].
The third puzzle is the
remarkably fast drop in the QSO population at redshifts
below about 2;
during this period the comoving emissivity of luminous
QSO's drops by orders of magnitude and with a halving time substantially
shorter than the concurrent cosmic expansion time scale [7].

%Of course, the details of the evolution of such complex and nonlinear
%astrophysical objects as galactic nuclei are sufficiently poorly
%understood to allow theorists enough flexibility to explain away
%the dramatic, if well known, QSO population evolution (Efstathiou \& Rees
%1988, Small \& Blandford 1992, Haehnelt \& Rees 1993).  However,
%it is perhaps worth recalling that the standard scenario would
%{\it predict} qualitatively different evolution if interpreted in a
%simple and straightforward way.  Namely, most structure formation
%scenarios (particularly in high $\Omega_o$ universes) predict that
%increasingly massive objects form at successively later epochs.
%Moreover, processes of gravitational collapse and accretion would be
%expected to accelerate and produce ever more massive and rapidly growing
%black holes, especially in the very high density environments at the
%cusps of the nuclei of bright galaxies.  This would lead to
%the expectation of a QSO population becoming more
%luminous and numerous with time as structure formation and the nonlinear
%evolution of galactic nuclei proceed.  Even if one postulates some limit
%to terminate the luminous QSO phase, such as exhaustion of the accretion fuel
%supply  or inability to tidally disrupt passing stars [8], i
%it is difficult to see
%why it should apply so synchronously;
%at a first approximation, all QSO activity is observed to end at the same
%cosmic time!

It is useful to recall that the natural {\it a priori
prediction} of the conventional model would be quite different.
Most structure formation
scenarios (particularly in high $\Omega_o$ universes) predict that
increasingly massive objects form at successively later epochs.
Moreover, processes of gravitational collapse and accretion would be
expected to accelerate and produce ever more massive and rapidly growing
black holes, especially in the very high density environments at the
cusps of the nuclei of bright galaxies.  This would lead to
the expectation of a QSO population becoming more
luminous and numerous with time as structure formation and the nonlinear
evolution of galactic nuclei proceed.  Even if one postulates some limit
to terminate the luminous QSO phase, such as exhaustion of the accretion fuel
supply  or inability to tidally disrupt passing stars [9],
it is difficult to see
why it should apply so synchronously as observed.
Of course, given our inability to reliably predict the details of the
complex, nonlinear evolution of galactic nuclei and black hole accretion
processes, it has proven possible to {\it a posteriori}
explain the QSO population evolution in various, sometimes {\it ad hoc},
ways [10-13].
%(Efstathiou \& Rees
%1988, Small \& Blandford 1992, Haehnelt \& Rees 1993).

Motivated by the Bahcall {\it et al.} [2-4] observations and these
shortcomings of the conventional model we here
investigate the alternative hypothesis that the Universe contains a
substantial population of massive ($\sim10^8M_\odot$) black holes
existing  independently of any host galaxy and perhaps even formed by
rather different physical mechanisms.  QSOs
are then identified with the accretion luminosity and other activity
generated when one of these ``naked" black holes collides with a galaxy
or a massive IGM cloud.
In addition to trying to account for the unexpected results of the
HST study, it is expected that such a scenario will be able to
naturally account for the large and extremely rapid decrease in QSO
activity in the recent history of the Universe.
Thus, although the postulated ``naked" black hole population may seem
rather {\it ad hoc}, it offers the possibility of understanding
at least two
otherwise quite puzzling observations.

This simple scenario raises a host of detailed theoretical considerations.
These include the efficiency with which the black hole can accrete
gaseous material
from the galaxy or cloud with which it is colliding (as a function of
gas density, angular momentum, composition, ...), the required number density
of the ``naked" black holes (relative to available baryons and other limits),
their formation mechanisms and
epochs, possible effects on the cosmic radiation background (CRB) spectrum,
gravitational
lensing consequences, relation to those QSOs and AGN which do
seem to reside in the nuclei of $L \sim L^*$ galaxies, possible explanations
for QSO phenomenologies and classifications, and so forth.
In this discussion, we consider a few of the more critical points briefly,
but we do not claim to have resolved any of these theoretical issues
conclusively.  Rather, our main purpose is to suggest a new qualitative
scenario for the nature of at least some QSO's and to show that it is
not immediately excluded by any simple considerations.

The typical black hole mass of interest is $\sim 10^8M_\odot$
at around $z\sim2$.  The characteristic luminosity is set by the
Eddington limit,
at which radiation pressure on free electrons balances gravitational forces:
$$L_E=4\pi GM_hm_p/\sigma_T=1.3\times 10^{46}M_8{\rm erg~ s}^{-1}
\eqno{(1)}$$
where $G$ is Newton's constant, $M_h$ is black hole mass,
$m_p$ is proton mass, $\sigma_T$ is the Thomson scattering
cross section and $M_8$ is the black hole mass in units of
$10^8M_\odot$.  This is a typical luminosity for bright
QSOs.  Accretion rates needed to maintain this luminosity are
of order $2\epsilon_{0.1}^{-1} M_\odot {\rm yr}^{-1}$, where $\epsilon_{0.1}$
is the fraction of the accreted material's rest energy which is emitted
radiatively in units of $0.1$.
In particular, it is expected that a black hole moving with a
characteristic velocity $v$ through a diffuse medium of density
$n$ will produce an accretion luminosity [14,15]
$$L = 1.0\times10^{45} M_8^2 \epsilon_{0.1} ({{v} \over {100 {\rm
km/s}}})^{-3}({{n} \over {3 {\rm cm^{-3}}}}) {\rm ergs/s}, \eqno{(2)}$$
%(Hunt 1971, Lacey and Ostriker 1985),
which is of order the luminosity of
the low redshift QSO's observed by Bahcall {\it et al.}.
Some combination of higher densities
in the ambient medium, lower encounter velocities and/or more massive black
holes would be required to achieve $10^{46}$ erg/s luminosities characteristic
of high redshift QSOs.

What cosmic density of black holes is required to produce the
observed QSO population?
Let us write the comoving number density of black holes as
$n_h$, and that of galaxies as $n_g$.  The encounter rate is
then written
$$R=n_hn_g(1+z)^6\sigma v,\eqno{(3)}$$
where $\sigma$ is the cross
section such that an encounter gives a QSO and $v$ the relative
velocity.  For an order-of-magnitude
estimate, let us assume that 30\% of baryons are captured in
galaxies
%of $10^8M_\odot$ mass
and 10\% becomes black holes of mass
$10^8M_\odot$. We assume that masses of galaxies are distributed
according to the Schechter luminosity function with $M_{\rm baryon}/L
\sim 10$, and that the relative velocities of a galaxy and a black hole
are typically
100km s$^{-1}$.
We also assume that QSO activity takes place when
black holes crosses the galaxy within its Holmberg radius
(at a baryonic surface density $\approx 10 M_\odot$ pc$^{-2}$); we estimate the
size of
a galaxy to be $\sim14(M_B/10^{11}M_\odot)^{0.4}$ kpc.
We take the slope of the Schechter function to be $\alpha=
-1.5$ in agreement with the numerous dwarfs
reported in several recent studies [16-18].
The integral over the Schechter function is dominated by
objects with $10^8-10^{10}M_\odot$ of gaseous material.

We note that the mass of $10^8M_\odot$ is
close to the minimum mass needed for bright QSOs; a
less massive galaxy would not supply sufficient fuel
to sustain the QSO luminosity through a characteristic
crossing time.
A fuel reservoir of $10^{9}$ $M_\odot$
could shine
for about $10^9$ years at  the Eddington luminosity
if $\epsilon_{0.1}=1$.
{}From (3) we find
that the encounter rate $R \sim 4.0\times10^{-22}(1+z)^6h^{-2}$ s$^{-1}$
Mpc$^{-3}$
or equivalently, $1.3\times10^{-4}(1+z)^6h^{-3}$ Mpc$^{-3}$ per inverse Hubble
time, where $h$ is Hubble's constant in units
of 100 km$^{-1}$ s$^{-1}$ Mpc$^{-1}$.
Thus, the cumulative number of collisions (which we identify as QSO
outbursts)
around $z\sim2$ is about
0.018$h^{-3}$ Mpc$^{-3}$ in physical density or 0.0007$h^{-3}$ Mpc$^{-3}$
in comoving units. These numbers are a conservative
estimate, since in reality,
we expect some correlation in the distribution of
QSOs and galaxies which would significantly increase the rate
of collisions.
In addition, QSO activity might result from the collision
of black holes with some of the denser clouds in the IGM, rather than
with a galaxy, thus further increasing the collision rate or
reducing the required black hole population.

We note that the total mass density of black holes that ever shone as QSOs
can be estimated reliably from their cumulative observed flux [19].
A modern estimate is $n_h \sim 0.001-0.002M_8^{-1}$
Mpc$^{-3}$ [20], in comoving coordinates.
The fact that this number density based on the observed QSO population
agrees with that inferred above from the collision rate calculation,
within the substantial uncertainties of the input parameters (including h)
and our very simplified treatment, is encouraging.
This density is about 0.1$h^{-3}$ that of luminous
galaxies and is at least $\sim 10^2h^{-3}$ times higher
than the peak number density of QSOs, which already suggests that
QSOs are made and fade one after another.

A particularly notable feature is the $(1+z)^6$ dependence of
the encounter rate.  Since we expect that the lifetime
of $10^8-10^9M_\odot$ QSOs is of the order of $0.6-6\times 10^8$ yr,
considerably shorter than the Hubble time, we predict that the number
density of QSOs decreases as $(1+z)^6$ towards $z=0$ in
gross qualitative agreement with observations.  Of course, since
the other factors in equation (3) ($n_g, \sigma, v$) might evolve with
redshift, the situation could be considerably more complex, but these
effects will be dominated by the $(1+z)^6$ factor unless the evolution
is extreme.

%Let us now examine whether the numbers assumed above is reasonable.
%The number of $10^8$ mass galaxies we assumed above corresponds to
%$10(1+z)^3$ Mpc$^{-3}$. At the first sight, this looks much more
%than the observed number of bright galaxies, 0.02Mpc$^{-3}$ for
%galaxies with $10^{10}L_\odot$.  If, however, the slope of
%Schechter luminosity function is as steep as -1.5, as recent
%observations of dwarf galaxies indicates, 1000 times less luminous
%galaxies are 50 times more numerous. If we consider the current
%belief that a large mass galaxies are the result of mergers of
%subluminous galaxies, this number may further be multipled by
%a factor of a few.  Therefore, the galaxy density we quoted is not
%an unreasonable number.  The number density of black holes is
%much less constrained, and our assumed density does not conflict
%with observations.

The Bahcall {\it et al.} observations [2-4] show that QSOs
are occasionally associated
with host galaxies, sometimes spirals and sometimes ellipticals.
This does not necessarily contradict the basic model presented here.  While
most QSOs would lose their activity rather quickly,
either due to the end of the collision or the exhaustion of fuel
in low mass galaxies,
close encounters with more massive galaxies will
sometimes lead to capture of the
black hole via dynamical friction and produce much longer lived
activity.
The dynamical
friction will eventually bring the black hole into the centre of a
galaxy.
The time scale for a $10^8 M_\odot$ black hole to spiral into the
nucleus of a typical giant galaxy from an initial radius of $\sim 10$
kpc is of order $10^{10}$ yrs [21].
It is interesting to see that the nucleus of 3C273 is
not at the centre, and there are a few other examples seen in the
QSO sample of Bahcall {\it et al}.  In a predictive sense, when the active
nucleus is found at  or near the centre of a galaxy, we would expect
a massive, high density host galaxy capable of producing strong
dynamical friction.

As for the formation of black holes, our suggestions are not more than
speculative. According to standard hierarchical clustering models, small
mass objects collapse earlier; typically one expects large numbers of bound
objects of $\sim 10^5-10^7M_\odot$ before $z\sim 10$.
Alternatively, more unconventional structure formation models such as
PBI [22]
or cosmic textures [23]
can produce very nonlinear structure
formation on small mass scales at early epochs.  Of course, the black hole
formation mechanism might be quite unrelated to those that form galaxies
and other familiar structures.  It is also worth noting that early formation
of a galaxy-independent population of massive black holes has been
invoked and investigated in a variety of other astrophysical connections
[15,24-28].

In any case, we
may suppose that a small
fraction of baryons go into black holes
when Compton cooling is very efficient (i.e., $z>10$), as a
generalization of the Rees \& Efstathiou scenario [9].
Some of such black holes may eventually grow to $10^8M_\odot$
by $z=3-4$. Let us assume that the initial black hole mass
is $M_{hi}$ formed at around $z\sim 10-20$. If a fraction $f$ of the binding
energy of the forming black holes is
deposited
into the CRB; then the amount of distortion to be observed
as Zel'dovich-Sunyaev effect is $y_c=\delta \rho/4 \rho_{\rm CRB}$,
which must be smaller than the observed limit $2.5\times10^{-5}$ [29].
This means that the initial
mass of black hole be smaller than $\sim 10^5M_\odot$ for $f=0.1$.
On the other hand,
the characteristic accretion time is $t_E=4\times 10^8$, and
the mass of a black hole can grow as fast as $M_h(t)\sim \exp(t/\epsilon t_E)$
where $\epsilon$ is the radiative efficiency, usually assumed
of the order of 0.1.  Hence, the available time is enough for
more than 20 e-folds, sufficient to bring the mass to $10^8M_\odot$
well before $z\sim 4$, though the availability of an adequate accretion
fuel supply is a nontrivial requirement [6].
This exponential increase of the hole mass,
and hence accretion luminosity, would explain the rapid rise
of a bright QSO population before $z=3-4$ [30].

Gravitational lensing provides a potential direct method for detecting
the postulated black hole population [31].  Unfortunately however, the
angular splittings, roughly 0.01 arc sec, and per source probabilities
of multiple imaging, less than $10^{-3}$ at $z=2$, are so small as to
preclude any useful tests of our hypothesis based on available data sets.

Our primary conclusions can then be stated as follows:  The conventional and
in many respects successful model for QSOs [1] is severely challenged by recent
HST data and has difficulty accounting for their well established
population evolution.  The alternative scenario suggested above could better
account for these observations and does not manifestly violate any other
empirical constraints.  Thus, it merits further exploration.

\bigskip
\bigskip
\centerline{\bf References}

\medskip
\singlespace

\noindent
1. Rees, M. J. 1984, ARA\&A, 22, 471.

\noindent
2. Bahcall, J. N., Kirhakos, S. \& Schneider, D. P. 1995, ApJ Lett., 435, L11.

\noindent
3. Bahcall, J. N., Kirhakos, S. \& Schneider, D. P. 1995, ApJ, in press.

\noindent
4. Bahcall, J. N., Kirhakos, S. \& Schneider, D. P. 1995, ApJ Lett., in press.

\noindent
5. Hartwick, F. D. A. \& Schade, D. 1990, ARA\&A, 28, 437.

\noindent
6. Turner, E. L., 1991, AJ, 101, 5.

\noindent
7. Turner, E. L. 1991, in {\it The Space Distribution of Quasars}, ed. D.
Crampton
(Ast. Soc. Pac., San Francisco), 361.

\noindent
8. Goodman, J. \& Lee, H. M. 1989, ApJ, 337, 84.

\noindent
9. Rees, M. J. 1990, Science, 247, 817.

\noindent
10. Efstathiou, G. \& Rees, M. J. 1988, MNRAS, 230, 5p.

\noindent
11. Small, T. A. \& Blandford, R. D. 1992, MNRAS, 259, 725.

\noindent
12. Haehnelt, M. G. \& Rees, M. J., 1993, MNRAS, 263, 168.

\noindent
13. Narayan, R. \& Yi, I. 1995, ApJ, in press.

\noindent
14. Hunt, R. 1971, MNRAS, 154, 141.

\noindent
15. Lacey, C. G. \& Ostriker, J. P. 1985, ApJ 299, 633.

\noindent
16. Impey, C., Bothum, G. \& Malin, D. 1988, ApJ, 330, 634.

\noindent
17. Turner, J. A., Phillips, S., Davies, J. I. \& Disney, M. J. 1993, MNRAS,
261, 39.

\noindent
18. Dalcanton, J. 1995, PhD Thesis (Princeton University).

\noindent
19. Soltan, A. 1982, MNRAS, 200, 115.

\noindent
20. Chokshi, A. \& Turner, E. L. 1992, MNRAS, 259, 421.

\noindent
21. Begelman, M. C., Blandford, R. D. \& Rees, M. J. 1980, Nature, 287, 307.

\noindent
22. Peebles, P. J. E. 1987, ApJ Lett., 315, L73.

\noindent
23. Gooding, A. K., Spergel, D. N. \& Turok, N. 1991, ApJ Lett, 372, L5.

\noindent
24. Carr, B. J. \& Rees, M. J. 1984, MNRAS, 206, 315.

\noindent
25. Gnedin, N. Yu. \& Ostriker, J. P. 1992, ApJ, 400, 1.

\noindent
26. Umemura, M., Loeb, A. \& Turner, E. L. 1993, ApJ, 419, 459.

\noindent
27. Loeb, A. \& Rasio, F. A. 1994, ApJ, 432, 52.

\noindent
28. Eisenstein, D. J. \& Loeb, A. 1995, ApJ, 443, 11.

\noindent
29. Mather, J. C. {\it et al.} 1994, ApJ, 420, 439.

\noindent
30. Schneider, D. P., Schmidt, M. \& Gunn, J. E. 1989, Astron J, 98, 1951.

\noindent
31. Press, W. H. \& Gunn, J. E. 1973, ApJ, 185, 397.

\vskip1cm

\noindent
ACKNOWLEDGEMENTS. We would like to thank John Bahcall and Sofia Kirhakos
for valuable discussions. This research was supported in part by US-NSF grants
PHY94-07194
at the Institute for Theoretical Physics in Santa Barbara and
AST94-19400 at Princeton
University.  MF also thanks the Fuji-Xerox Corporation for generous
support.

\end